\documentclass[graybox]{svmult}

\usepackage{mathptmx}       
\usepackage{helvet}         
\usepackage{courier}        
\usepackage{type1cm}        
\usepackage{makeidx}         
\usepackage{graphicx}        
\usepackage{psfrag,multirow}
\usepackage{mathbbol}
\usepackage{multicol}        
\usepackage[bottom]{footmisc}

\makeindex             

\begin{document}

\title*{Kinetic exchange opinion model: solution in the single parameter map limit}
\author{Krishanu Roy Chowdhury, Asim Ghosh, Soumyajyoti Biswas and Bikas K. Chakrabarti}
\institute{Krishanu Roy CHowdhury \at Saha Institute of Nuclear Physics, 1/AF, Bidhannagar, Kolkata-700064, \email{krishanu.1987@gmail.com}
\and Asim Ghosh \at Saha Institute of Nuclear Physics, 1/AF, Bidhannagar, Kolkata-700064, \email{asim.ghosh@saha.ac.in} 
\and Soumyajyoti Biswas \at Saha Institute of Nuclear Physics, 1/AF, Bidhannagar, Kolkata-700064, \email{soumyajyoti.biswas@saha.ac.in}
\and Bikas K. Chakrabarti \at Saha Institute of Nuclear Physics, 1/AF, Bidhannagar, Kolkata-700064, \email {bikask.chakrabarti@saha.ac.in}}
%
%
\maketitle
\begin{abstract}
\noindent We study a recently proposed kinetic exchange opinion model (Lallouache et. al., Phys. Rev E 82:056112, 2010) 
in the limit of a single parameter map. Although it does not include the essentially complex behavior of the
 multiagent version, it provides us with the insight regarding the choice of order parameter for the system as 
well as some of its other dynamical properties. We also study the generalized two-parameter version of the model,
 and provide the exact phase diagram. The universal behavior along this phase boundary in terms of the suitably
 defined order parameter is seen.
\end{abstract}

\section{Introduction}\label{sec:1}
\noindent Dynamics of opinion and subsequent emergence of consensus in a society are being extensively studied recently \cite{ESTP,Stauffer:2009,
Castellano:RMP,Galam:1982,Liggett:1999, Sznajd:2000,Galam:2008,Sen:opin}.
Due to the involvement of many individuals, this type of dynamics  in a society can be treated as an example of
a complex system, thus enabling the use of conventional tools of statistical mechanics to model it \cite{Hegselman:2002,Deffuant:2000,Fortunato:2005,
forunato,Toscani:2006}. Of course,
it is not possible to capture all the diversities of human interaction through any model of this kind. But often it is our interest
to find out the global perspectives of a social system, like average opinion of all the individuals regarding an issue, where the 
intricacies of the interactions, in some sense, are averaged out. This is similar to the approach of kinetic theory, where 
the individual atoms, although following a deterministic dynamics, are treated as randomly moving objects and the macroscopic 
behaviors of the whole system are rather accurately predicted. 

Indeed, there have been several attempts to realize the human interactions in terms of kinetic exchange of opinions between 
individuals \cite{Hegselman:2002,Deffuant:2000,Toscani:2006}. Of course, there is no conservations in terms of opinion. Otherwise, this is similar to momentum exchange
between the molecules of an ideal gas. These models were often studied using a finite confidence level, i.e., agents having
opinions close to one another interact. However, in a recently proposed model \cite{Lallouache:2010},  unrestricted interactions between all the agents
were considered. The single parameter in the model described the `conviction'  which is a measure of an agent's tendency to retain
his opinion and also to convince others to take his opinion. It was found that beyond a certain value of this `conviction parameter'
the `society', made up of $N$ such agents, reaches a consensus, where majority shares similar opinion. As the opinion values could
take any values between [-1:+1], a consensus means a spontaneous breaking of a discrete symmetry.

There have been subsequent studies to generalize this model, where the `conviction parameter' and `ability to influence' were taken as
independent parameters \cite{Sen:2010}. In that two-parameter version, similar phase transitions were observed. However, the critical behaviors
in terms of the usual order parameter, the average opinion, were found to be non-universal. There have been other extensions in terms of  a 
phase transition induced by negative interactions \cite{bcs}, an exact solution in a 
discrete limit \cite{sb}, the effect of non-uniform conviction and update rules in these discrete variants \cite{nuno}, 
a generalized map version \cite{asc:2010}, a percolation transitions in a square lattices \cite{akc} and the effect of bounded 
confidence \cite{ps} in these models.

In the present study, we investigate the single parameter map version of the model, also proposed in Ref.\cite{Lallouache:2010}. Although
the original model is difficult to tackle analytically, in this mean field limit, it can simply be conceived as a random walk. Using standard random 
walk statistics, several static and dynamical quantities have been calculated. We show that the fraction of extreme opinion behaves like 
the actual order parameter for the system, and the average opinion shows unusual behavior near critical point. The critical behavior of 
the order parameter and its relaxation  behavior near and at the critical 
point have been obtained analytically which agree with  numerical simulations.

\section{Model and its map version}

\noindent Let the opinion of any individual ($i$) at any time ($t$) is represented by a real valued variable $O_i(t)$ ($-1\leq O_i<1$). The kinetic exchange model of opinion pictures the opinion exchange between two agents like a scattering process in 
an ideal gas. However, unlike ideal gas, there is no conservation of the total opinion. This is similar to the kinetic exchange
models of wealth redistribution, where of course the conservation was also present \cite{CC-CCM}. The (discrete time) exchange equations of the model read
\begin{equation}
O_i(t+1)=\lambda O_i(t)+\lambda \epsilon O_j(t),
\label{exchange}
\end{equation}  
and a similar equation for $O_j(t+1)$, where $O_i(t)$ is the opinion value of the $i$-th agent at time $t$, $\lambda$ is the 
`conviction parameter' (considered to be equal for all agents for simplicity) and $\epsilon$ is an annealed  random number 
drawn from a uniform and continuous distribution between [0:1], which is the probability with which $i$ and $j$ interact (see \cite{Lallouache:2010}). 
Note that the choice of $i$ and $j$ are unrestricted, making the effective interaction range to be infinite. The opinion 
values allowed are bounded between the limits $-1\le O_i(t)\le +1$. So, whenever the opinion values are predicted to be greater 
(less) than +1 (-1) following Eq.~(\ref{exchange}), it is kept at +1 (-1). This bound, along with Eq.~(\ref{exchange}), defines the
dynamics of the model.

This model shows a symmetry breaking transition at a critical value of $\lambda$ ($\lambda_c\approx 2/3$). The critical
behaviors were studied using the average opinion  ${O_a}=|\sum\limits_{i=1}^NO_i(t\to \infty)|/N$ \cite{bcc}. An alternative
parameter was also defined in Ref.\cite{Lallouache:2010}, which is the fraction of agents having extreme opinion values. This quantity also
showed critical behaviors at the same transition point.

The model in its original form is rather difficult to tackle analytically (it can be solved in some special limits though \cite{sb}).
 However, as it is a fully connected model, a mean field approach would lead to the following evolution equation for the
single  parameter opinion value (cf. \cite{Lallouache:2010})
\begin{equation}
\label{map}
O(t+1)=\lambda(1+\epsilon)O(t).
\end{equation}
This is, in fact, a stochastic map with the bound $|O(t)|\le 1 $. For all subsequent discussions, whenever an explicit time dependence of a 
quantity is
not mentioned, it denotes the steady state value of that quantity and a subscript $a$ denotes the average over the randomness (i.e., ensemble average).
 As we will see from the subsequent discussions, this
map can be conceived as a random walk with a reflecting boundary.  As in the case of the multiagent version, the distribution 
of $\epsilon$ does not play any role in the critical behavior. We have considered two distributions, one is continuous
in the interval [0:1] and the other is 0 and 1 with equal probability. Both of these give similar critical behavior.

We also briefly discuss the two-parameter model, where the `conviction' of an agent and the ability to convince others were
taken as two independent parameters \cite{Sen:2010}. In that context, the map would read
\begin{equation}
O(t+1)=(\lambda+\mu\epsilon)O(t),
\label{mapp}
\end{equation}
where $\mu$ is the parameter determining an agent's ability to influence others. As before, $|O(t)|\le 1$.

\section{Results}
\subsection{Random walk picture}
\noindent One can study the stochastic map  in Eq.~(\ref{map}) by describing it in terms of random walks.
Writing $X(t)=\log({O(t)})$ (for all subsequent discussions we always take $O(t)$ to be positive), Eq.~(\ref{map}) can be written as
\begin{equation}
X(t+1)=X(t)+\eta,
\label{rw}
\end{equation}
where, $\eta(t)=\log[\lambda(1+\epsilon)]$. As is clear from the above equation, it actually describes a random walk 
with a reflecting boundary at $X=0$ to take the upper cut-off of $O(t)$ into account. Depending upon the value of
$\lambda$, the walk can be biased to either ways and is unbiased just at the critical point. As one can   average independently 
over these additive terms in Eq.~(\ref{rw}), this gives an easy way
to estimate the critical point \cite{Lallouache:2010}. An unbiased random walk would imply $\langle \eta\rangle=0$ i.e.,
\begin{equation}
\int\limits_0^1\log[\lambda_c(1+\epsilon)] d\epsilon=0
\end{equation} 
giving $\lambda_c=e/4$, where we have considered an uniform distribution of $\epsilon$ in the limit [0:1]. This estimate
matches very well with numerical results of this and earlier works \cite{Lallouache:2010}.
\begin{figure}[tb]
\centering \includegraphics[width=9cm]{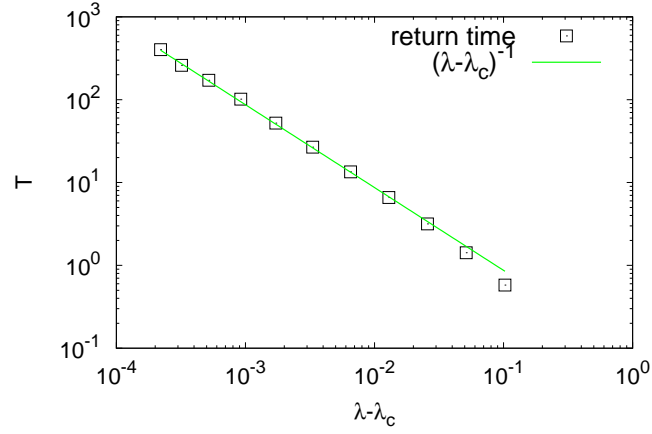}
   \caption{The average return time $T$ of ${O}(t)$ to 1 in the map described in Eq.~(\ref{map}) is plotted with $(\lambda-\lambda_c)$. It 
shows a divergence with exponent 1 as is predicted from Eq.~(\ref{returntime})}
\label{time-map-conti}
\end{figure}
\begin{figure}[tb]
\centering \includegraphics[width=9cm]{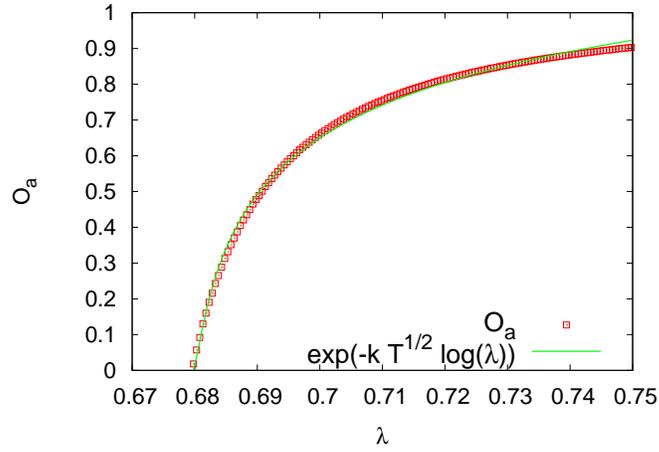}
   \caption{ $O_a$ is plotted with $\lambda$. The data points are results of numerical simulations, which fits rather well
with the solid line predicted from Eq.~(\ref{averageO}), with $k=0.7$}
\label{order-map-conti}
\end{figure}
In order to guess the $\lambda$ dependence of ${O_a}$ in the ordered region, we  first  estimate the ``average  return time'' 
$T$ (return time is the time between two successive reflections from $X=0$) as a function of bias of the walk. For this uniform 
distribution of $\epsilon$, the average position to which the walker goes following a reflection from the barrier is
$(\lambda+1)/2$. The average amount of contribution in each step is given by $\int\limits_o^1\log[\lambda(1+\epsilon)]d\epsilon=\log(\lambda/\lambda_c)$.
This, in fact, is a measure of the bias of the walk, which vanishes linearly with $(\lambda-\lambda_c)$ as $\lambda\to \lambda_c$.
So, in this map picture, one would expect that on average by multiplying this  $\lambda/\lambda_c$  factor $T$  times (i.e., adding 
$\log(\lambda/\lambda_c)$ $T$ times in the random walk picture), $O(t)$ would reach 1 from $(\lambda+1)/2$. 
Therefore,
\begin{equation}
\frac{\lambda +1}{2}\left( \frac{\lambda}{\lambda_c} \right)^T=1,
\end{equation}
giving
\begin{equation}
T=- \frac{\log\lambda}{\log\lambda-\log\lambda_c}  \approx - \frac{\log\lambda}{\lambda-\lambda_c}  
\label{returntime}
\end{equation}
for $\lambda\to \lambda_c$.  Clearly, the average return time diverges near the critical point obeying a power law: 
$T \sim (\lambda-\lambda_c)^{-1}$. In Fig.~\ref{time-map-conti} we have plotted this average return time as a function of $\lambda$. The
power-law divergence agrees very well with the prediction. 

The steady state average value of $X(t)$ i.e., $X_a$ (and correspondingly $O_a$) is expected to be proportional to $\sqrt{T}$ in steps of $\log\lambda$:
\begin{equation}
{X_a}\sim \sqrt{T}\log\lambda = k\sqrt{T}\log\lambda,
\end{equation}
where $k$ is a constant. This
gives 
\begin{equation}
{O_a}= \exp[-{k}|\log\lambda|^{3/2} (\lambda-\lambda_c)^{-1/2}].
\label{averageO}
\end{equation}

The above functional form fits quite  well (see Fig.~\ref{order-map-conti}) with the numerical simulation results near  the critical point.
It may be noted that the numerical results for the kinetic opinion exchange Eq.~(\ref{map}) also fits quite well with this expression (Eq.~(\ref{averageO})).
\begin{figure}[tb]
\centering \includegraphics[width=9cm]{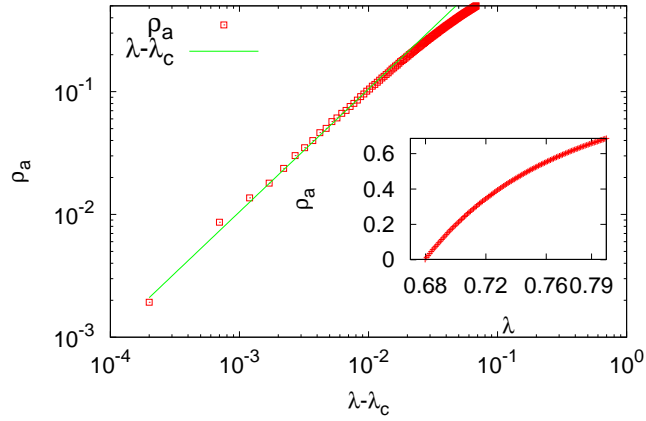}
   \caption{The average condensation fraction (probability that $O=1$) $\rho_a$ is plotted with $(\lambda-\lambda_c)$. A linear fit in the log-log scale gives
the growth exponent 1, as predicted from Eq.~(\ref{condf}). Inset shows the variation of $\rho$ with $\lambda$.}
\label{one-map-conti}
\end{figure}
We note that ${O_a}$ increases from zero at the critical point and eventually reaches 1 at $\lambda=1$. But
its behavior close to critical point cannot be fitted with a power-law growth usually observed for order-parameters.
Such peculiarity in the critical behavior of ${O_a}$ compels us to exclude it as an order 
parameter though it satisfies some other good qualities of an order parameter. Instead, we  consider the average
`condensation fraction' $\rho_a$ as the order parameter. In the multi-agent version, it was defined as the fraction of agents
having extreme opinion values i.e., -1 or +1. In this case it is defined as the probability that $O(t)=1$.
We denote this quantity by $\rho (t)$. As is clear from the definition, one must have
\begin{equation}
\rho_a\sim \frac{1}{T},
\label{condf}
\end{equation} 
where, $T$ is the return time of the walker. As $T\sim (\lambda-\lambda_c)^{-1}$, $\rho_a\sim (\lambda-\lambda_c)^{\beta} $ with $\beta=1$.
This behavior is clearly seen in the numerical simulations (see Fig.~\ref{one-map-conti}). 

\begin{figure}[tb]
\centering \includegraphics[width=9cm]{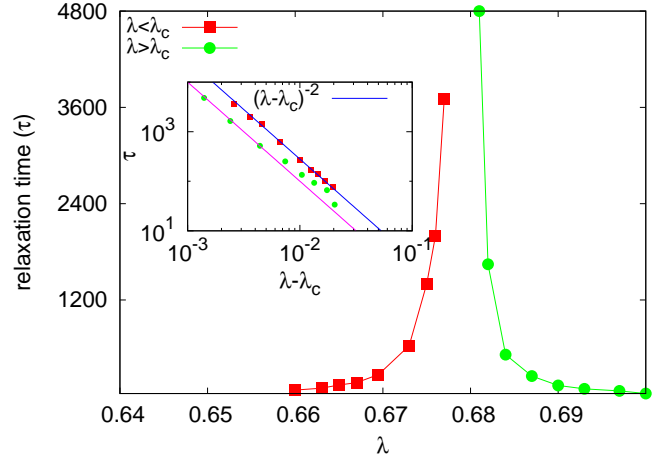}
   \caption{The average relaxation time for ${O_a}(t)$ is plotted with $\lambda$. This shows a prominent divergence as the critical point is approached.
In the inset, the relaxation time is is plotted in the log-log scale against $(\lambda-\lambda_c)$. The exponent is 2 as is expected from Eq.~(\ref{relaxtime})}
\label{relax-map-conti}
\end{figure}

Also, the relaxation time shows a divergence as the critical point is approached. We argue that there is a single relaxation time scale
for both $O(t)$ and $\rho (t)$. So we calculate the divergence of relaxation time for $O(t)$ and numerically show that the results agree
very well with the relaxation time divergence for both $O(t)$ and $\rho (t)$. Consider the subcritical regime where the random walker is
biased away from the reflector (at the origin) and would have a probability distribution for the position of the walker as
\begin{equation}
p(X)=\frac{A}{\sqrt{t}}\exp[-B(X-vt)^2)/t],
\end{equation}
 where $v\sim 1/T\sim (\lambda-\lambda_c)$ is the net bias and  constants $A$, $B$ do not depend on $t$. One can  therefore obtain the probability distribution  $P$ of $O$ using $p(X)dX=P(O)dO$,
\begin{equation}
P(O)= \frac{A}{\sqrt{t}}\frac{1}{O}\exp[-B(\log{O}-vt)^{2}/{t}].
\end{equation}
Hence
\begin{eqnarray}
{O_a}(t) &=& \int_{0}^{1}OP(O)dO ,
\nonumber \\
&=& \frac{A}{\sqrt{t}}\int_{0}^{1}\exp[-B(\log{O}-vt)^{2}/{t}]dO
\nonumber \\
&\sim& \frac{A}{\sqrt{t}}\exp(-Bv^{2}{t}),
\label{relaxtime}
\end{eqnarray}
in the long time limit, giving a time scale of relaxation $\tau\sim v^{-2} \sim (\lambda-\lambda_c)^{-2}$.
We have fitted the relaxation of ${O_a}(t)$, obtained numerically, with an exponential decay and found $\tau$. As can be observed from Fig.~\ref{relax-map-conti} it shows
a clear divergence close to critical point with exponent 2.
\begin{figure}[tb]
\centering \includegraphics[width=9cm]{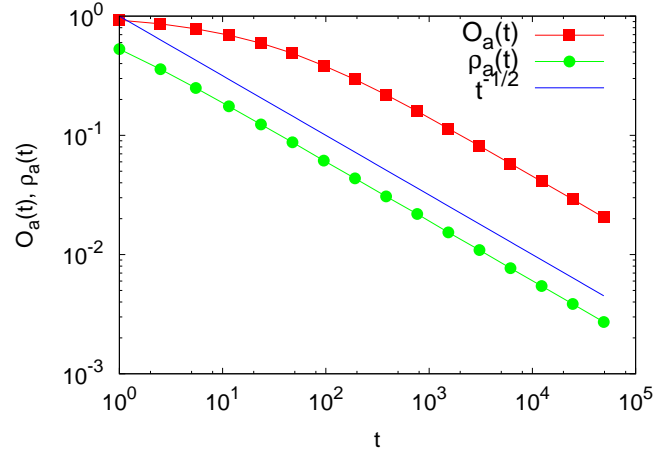}
   \caption{The time dependence of both ${O_a}(t)$ and $\rho (t)$ are plotted at the critical point in the log-log plot.
The linear fit shows a time variation of the form $t^{-\delta}$ with $\delta=1$, as is expected from Eq.~(\ref{relaxtime}).}
\label{delta-map-conti}
\end{figure}

 We have  obtained the relaxation time of ${\rho_a}(t)$ and it also shows similar divergence.
Note that at $\lambda=\lambda_c$, $v=0$ and it follows from Eq.~(\ref{relaxtime}) that $O_a(t)\sim t^{-1/2}$. This behavior is also confirmed numerically
(see Fig.~\ref{delta-map-conti}). The average condensation fraction $\rho_a(t)$ too follows this scaling, giving $\delta=1/2$ (as order parameter relaxes
 as $t^{-\delta}$ at critical point).

\begin{figure}[tb]
\centering \includegraphics[width=9cm]{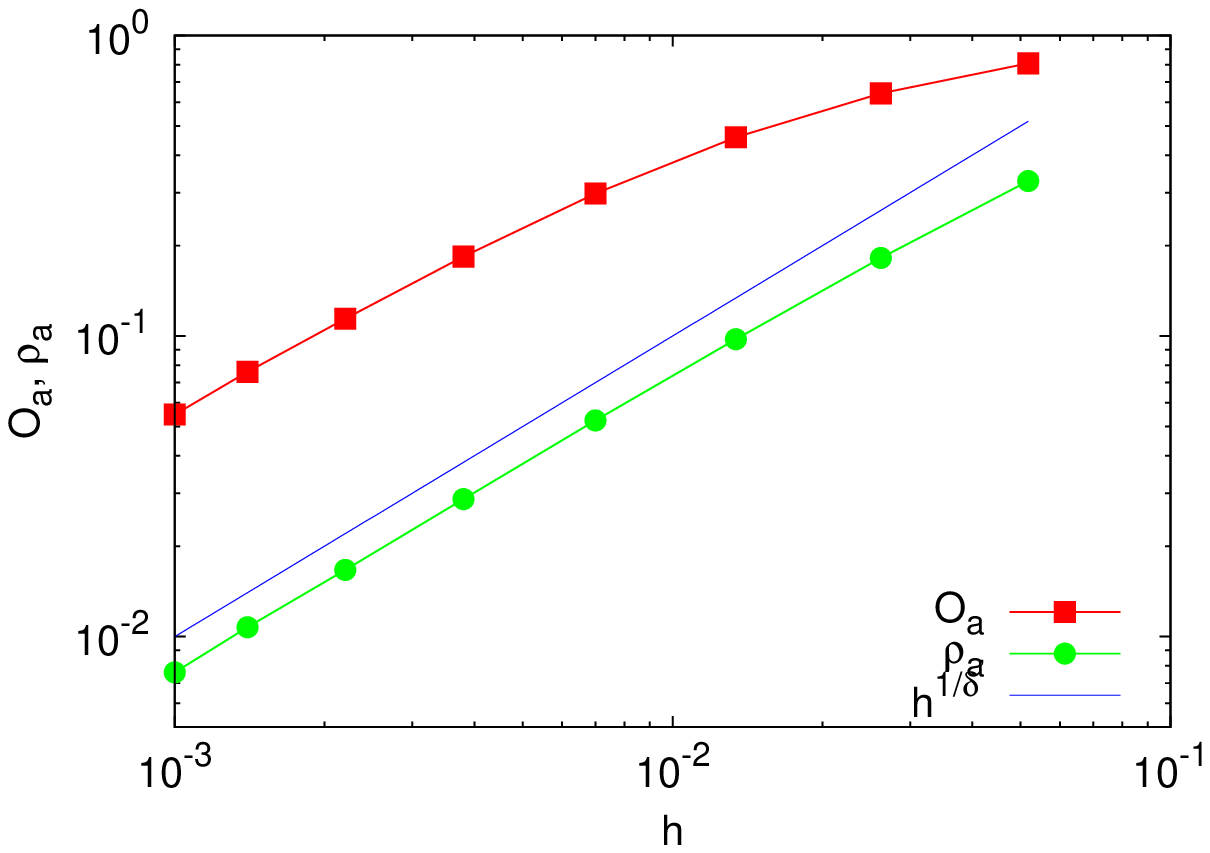}
   \caption{The variation of $O_a$ and $\rho_a$ are plotted against the external field $h$ at the critical point $\lambda=\lambda_C$. The linear fit in log-log scale shows $\delta^{\prime}=1$.}
\label{field-map-conti}
\end{figure}
We have also investigated the effect of having an external field linearly coupled with $O(t)$. In the multiagent scenario, this can have
the interpretation of the influence of media. The map equation now reads,
\begin{equation}
O(t+1)=\lambda(1+\epsilon_t)O(t)+h O(t),
\end{equation}
where $h$ is the field (constant in time). We have studied the response of ${O_a}$ and ${\rho_a}$ at $\lambda=\lambda_c$ due to application of 
small $h$. We find that (see Fig.~{\ref{field-map-conti}}) both grows linearly with $h$. One expects the order parameter to scale with external field at the critical 
point as ${\rho_a}\sim h^{1/\delta^{\prime}}$. In this case $\delta^{\prime}=1$. 

\subsection{Random walk with discrete step size}
\noindent One can simplify the random walk mentioned above and make it a random walk with discrete step sizes. This can be done by
considering the distribution of $\epsilon$ to be a double delta function, i.e., $\epsilon=1$ or $0$ with equal probability. This will
make $\eta(t)$ in Eq.~(\ref{map}) to be $\log\lambda$ or $\log(2\lambda)$ with equal probability. Below critical point, both steps
are in negative direction (away from reflector) and consequently taking the walker to $-\infty$. Exactly at critical point 
($\lambda=\lambda_c$) the step sizes become equal and opposite i.e., $\log\lambda_c=-\log(2\lambda_c)$ giving $\lambda_c= 1/\sqrt2$.  
Above critical point, one of the steps 
is positive and the other is negative. However, the magnitudes of the steps are different. This unbiased walker (probability of taking positive
and negative steps are equal) with different step sizes can  approximately be mapped to a biased walker with equal step size in both directions. 
To do that consider the probability $p(x,t)$ that the walker is at position $x$ at time $t$. One can then write the master equation
\begin{equation}
p(x,t+1)=\frac{1}{2}p(x+a,t)+\frac{1}{2}p(x+a+b,t),
\end{equation}
where $a=\log \lambda$ and $b=\log 2$. Clearly,
\begin{equation}
\frac{\partial p(x,t)}{\partial t}=\left(a+\frac{b}{2}\right)\frac{\partial p(x,t)}{\partial x}+\left(\frac{a^2}{2}+\frac{ab}{2}+\frac{b^2}{4}\right)\frac{\partial^2p(x,t)}{\partial x^2}.
\label{diff1}
\end{equation}
Now the master equation for the usual biased random walker can be written as 
\begin{equation}
 p(x,t+1)= p^{\prime}p(x+a^{\prime},t)+q^{\prime}p(x-a^{\prime},t),
\end{equation}
where $p^{\prime}$ and $q^{\prime}$ denote respectively the probabilities of taking  positive and negative steps ($p^{\prime}+q^{\prime}=1$) and $a^{\prime}$ is the (equal) step size in either direction. The differential form of this equation reads 
\begin{equation}
\frac{\partial p(x,t)}{\partial t}=(p^{\prime}-q^{\prime})a^{\prime}\frac{\partial p(x,t)}{\partial x}+\frac{a^{\prime2}}{2}\frac{\partial^2p(x,t)}{\partial x^2}.
\label{diff2}
\end{equation}

\begin{figure}[tb]
\centering \includegraphics[width=9cm]{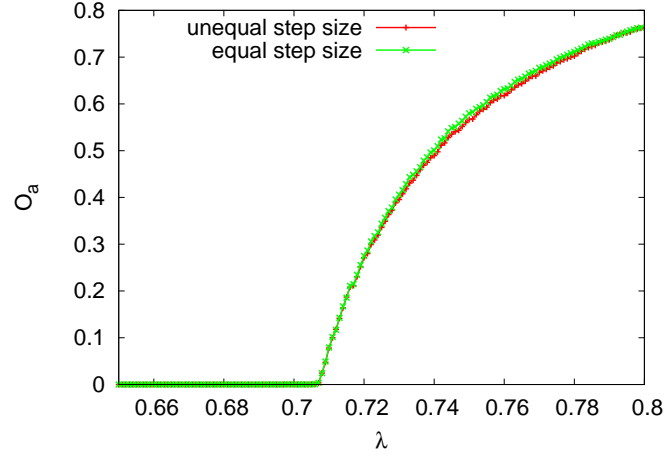}
   \caption{The comparison of the biased walk with equal step size with the original walk is shown. Reasonable agreement is seen for a 
wide range of $\lambda$ values.}
\label{comp-map-conti}
\end{figure}

Comparing these  equations (\ref{diff1}) and (\ref{diff2}),  one gets
\begin{eqnarray}
p^{\prime} &=& \frac{1}{2}\left[1+\frac{a+b/2}{a^{\prime}}\right] \nonumber \\
q^{\prime} &=& \frac{1}{2}\left[1-\frac{a+b/2}{a^{\prime}}\right] \nonumber \\
a^{\prime}&=& -\sqrt{(\log(\lambda)-\log(\lambda_c))^2+(\log(\lambda_c))^2}.
\end{eqnarray}
Therefore, as $\lambda\to \lambda_c$, the bias $(p^{\prime}-q^{\prime})\sim (\lambda-\lambda_c)/a^{\prime}$.
These are consistent with the earlier calculations where we have taken the bias to be proportional to $(\lambda-\lambda_c)$.
To check if this mapping indeed works, we have simulated a biased random walk with above mentioned parameters and found it
to  agree with the original walk (see Fig.~\ref{comp-map-conti}).
Similar to the approach taken for the continuous step-size walk, one can find the return time of the walker. This time the walker
is exactly located at $\lambda$ after it is reflected from the barrier. The return time again diverges as $(\lambda-\lambda_c)^{-1}$. 
Also ${O_a}(t)$ will have a similar form upto some prefactors. Condensation fraction will increase linearly with $(\lambda-\lambda_c)$
close to the critical point. All the other exponents regarding the relaxation time, time dependence at the critical point and dependence with 
external field are same as before. This shows that the critical behavior is universal with respect to changes in the distribution of $\epsilon$.

\subsection{Two parameter map}
\noindent As the multi-agent model was generalized in a two-parameter model \cite{Sen:2010}, one can also study the map version of that two-parameter model.
It would read
\begin{equation}
{O}(t+1)=(\lambda+\mu\epsilon){O}(t).
\label{map2} 
\end{equation}
As before, one can take $\log$ of both sides and in similar notations
\begin{equation}
X(t+1)=X(t)+\log(\lambda+\mu\epsilon).
\end{equation}
This can also be seen as a biased random walk. In fact, one can write the above equation as
\begin{equation}
X(t+1)=X(t)+\log[\lambda(1+\epsilon^{\prime})],
\end{equation}
where $\epsilon^{\prime}=(\mu/\lambda)\epsilon$. This effectively changes the limit of the distribution of the stochastic parameter. Here
$\epsilon^{\prime}$ is distributed between $0$ and $\mu/\lambda$. One can then do the earlier exercise with this version as well.
If one makes $\epsilon$ discrete, this is again a walk with unequal step sizes, which can again
be mapped to a biased walk with equal step sizes. Therefore, in either case, some pre-factors will  change, but the critical behavior will 
be the same as before. The critical behavior is, therefore, universal when studied in terms of the proper order parameter (condensation fraction).
\begin{figure}[tb]
\centering \includegraphics[width=9cm]{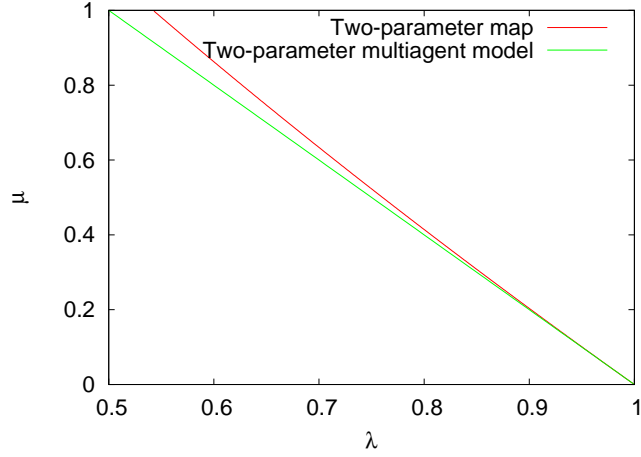}
   \caption{The phase diagram for the two-parameter map as predicted from Eq.~(\ref{phdia}) (upper line). The phase diagram for the multi-agent version ($\mu_c=2(1-\lambda_c)$) is also plotted for comparison (lower line).}
\label{ph-dia}
\end{figure}
For uniformly distributed $\epsilon$ (in the range [0:1]), one can get the expression for the phase boundary from the equation
\begin{equation}
\int\limits_0^1\log(\lambda+\mu\epsilon)d\epsilon=0,
\end{equation}
which gives
\begin{equation}
\log(\lambda_c+\mu_c)+\frac{\lambda_c}{\mu_c}\log\left(\frac{\lambda_c+\mu_c}{\lambda_c}\right)=1.
\label{phdia}
\end{equation}
Of course, this gives back the $\lambda_c=e/4$ limit when $\lambda=\mu$. The phase boundary is plotted in Fig.~\ref{ph-dia}. It also agrees with
numerical simulations.

\subsection{A map with a natural bound}
\noindent In the maps mentioned above, the upper (and lower) bounds
are additionally provided with the evolution dynamics. Here we study
a map where this bound occurs naturally. We intend to see its effect
on the dynamics.

Consider the following simple map
\begin{equation}
O(t+1)=\tanh[\lambda(1+\epsilon)O(t)].
\label{tanh}
\end{equation}
Due to the property of the function, the bounds in the values of $O(t)$
are specified within this equation itself. This map shows a spontaneous
symmetry breaking transition as before. This function appears in mean field
treatment of Ising model, but of course without the stochastic parameter.

\begin{figure}[tb]
\centering \includegraphics[width=9cm]{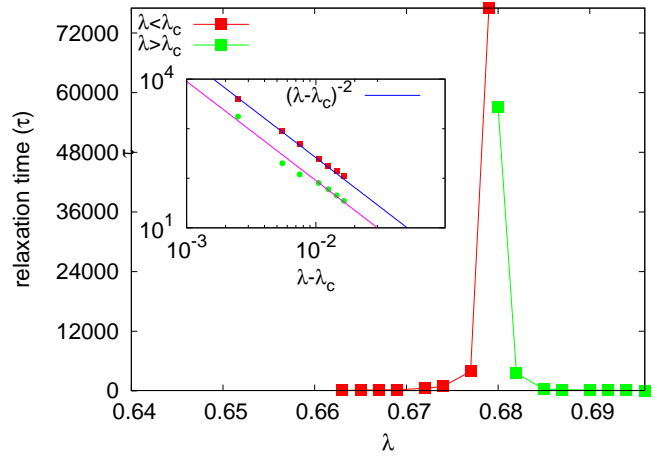}
   \caption{The divergence of relaxation time is shown for the map described by Eq.~(\ref{tanh}) for both sides of the critical point $\lambda_c=e/4$.
The inset shows that the exponent is 2 as is argued in the text.}
\label{tanh-map}
\end{figure}
Numerical simulations show that the $O_a$ behaves as $(\lambda-\lambda_c)^{1/2}$.
An analytical estimate of $\lambda_c$ can be made by linearizing the map for small
values of $O(t)$, which is valid only at critical point. Of course, after linearization the map is the same as the initial single parameter map, 
giving $\lambda_c=e/4$. The relaxation of $O_a(t)$ at $\lambda=\lambda_c$ behaves as $t^{-1/2}$ as before. These are also seen from numerical simulations. 
However, apart from the critical point, the map is strictly non-linear. Hence the results for its linearized version do not hold except for critical point. Also $\rho=0$ here always.

To check the divergence of the relaxation time at the critical point, it is seen that it follows $\tau\sim (\lambda-\lambda_c)^{-2}$. Of course, 
in the deterministic version (mean-field Ising model), the exponent is $1$. But as this map has stochasticity, the time exponent is 
doubled (see Fig.~\ref{tanh-map}) (i.e., the relaxation time is squared
as happens for  random motion as opposed to  ballistic motion).
\section{Summary and conclusions}

\noindent In this paper we have studied the simplified map version of a recently proposed opinion dynamics model.
The single parameter map was proposed in Ref.\cite{Lallouache:2010} and the critical point was estimated. Here we
study the critical behavior in details as well as propose the two-parameter map motivated from \cite{Sen:2010}. The
phase diagram is calculated exactly. 

The maps can be cast in a random walk picture with reflecting boundary. Then using the standard random walk 
statistics, some steady state as well as dynamical behaviors are calculated and these are compared with the corresponding numerical results. It is observed that the usual
 parameter of the system, i.e. the average value of the  opinion ($O_a$) in the steady state, does not follow any power-law scaling (see Eq.~(\ref{averageO})). 
In fact, it is the condensation fraction, or in this case the probability ($\rho_a$) that the opinion values touch the
limiting value, turns out to be  the proper order parameter, showing a power-law scaling behavior near critical point. 

The dynamical behaviors of the two quantities ($O_a(t)$ and $\rho_a(t)$) are similar. We have calculated the power-law relaxation of these
quantities at the critical point, which compares well with simulations. Also, the divergence of the relaxation time on both sides of criticality
shows similar behavior. We have also studied the effect of an ``external field'' (representing media or similar external effects) in these models. At critical point,
both the quantities grow linearly with the applied field. The average fluctuation in both of these quantities 
show a maximum near the critical point. These theoretical behavior fits well with the numerical results.

In summary, we develop an approximate mean field theory for the dynamical phase transition  observed for the map Eq.~(\ref{map}) and find the average 
condensation fraction $\rho_a\sim (\lambda-\lambda_c)^{\beta}$ with $\beta=1$ behave as the order parameter for the transition and it has typical 
relaxation time $\tau\sim (\lambda-\lambda_c)^{-z}$ with $z=2$ and at  critical point $\lambda=\lambda_c$($=e/4$) decays as $t^{-\delta}$ with $\delta=1/2$.

\bibliographystyle{elsarticle-num}
\bibliography{<your-bib-database>}


\end{document}